\documentclass[12pt]{article}
\usepackage{amssymb}
\usepackage{amsmath}
\usepackage{amscd}
\usepackage{latexsym}

\catcode`@=11
\renewcommand{\section}{\@startsection{section}{1}{0pt}{\medskipamount}
{\medskipamount}{\large\bf}}
\numberwithin{equation}{section}
\catcode`@=12

\def\a{\alpha}
\def\b{\beta}
\def\g{\gamma}
\def\D{\Delta}

\def\ve{\varepsilon}

\def\th{\theta}

\def\o{\omega}
\def\Om{\Omega}

\newcommand{\R}{{\mathbb R}}

\newcommand{\sfrac}[2]{{\textstyle\frac{#1}{#2}}}
\newcommand{\half}{{\sfrac12}}
\renewcommand{\>}{{\rangle}}

\newcommand{\+}{{\dagger}}
\newcommand{\pa}{{\partial}}
\newcommand{\diff}{{\mathrm{d}}}

\newcommand{\im}{{\mathrm{i}}}
\newcommand{\ep}{{\mathrm{e}}}

\newcommand{\tiH}{\widetilde{H}}
\newcommand{\tiL}{\widetilde{L}}
\newcommand{\tix}{\widetilde{x}}
\newcommand{\tip}{\widetilde{p}}
\newcommand{\tir}{\widetilde{r}}
\newcommand{\tipsi}{\widetilde{\Psi}}
\newcommand{\tiu}{\widetilde{u}}
\newcommand{\tiE}{\widetilde{E}}
\newcommand{\tiv}{\widetilde{v}}
\newcommand{\tie}{\widetilde{\ve}}
\newcommand{\tih}{\widetilde{h}}
\newcommand{\tiq}{\widetilde{q}}

\newcommand{\cD}{{\cal{D}}}

\newcommand{\with}{{\quad{\rm with}\quad}}
\newcommand{\for}{{\quad{\rm for}\quad}}
\renewcommand{\and}{{\quad{\rm and}\quad}}
\newcommand{\und}{{\qquad{\rm and}\qquad}}
\renewcommand{\=}{\ =\ }

\newcommand{\beq}{\begin{equation}}
\newcommand{\eeq}{\end{equation}}
\newcommand{\bea}{\begin{eqnarray}}
\newcommand{\eea}{\end{eqnarray}}

\advance \textheight by \topskip
\makeatletter
\@addtoreset{equation}{section}

\makeatother

\begin{document}
\begin{titlepage}
\setcounter{page}{0}
\begin{flushright}
ITP--UH--11/13\\
CCNY-HEP-13/03
\end{flushright}

\vskip 1.5cm

\begin{center}

{\LARGE\bf
The quantum angular Calogero--Moser model
}

\vspace{12mm}
{\Large Mikhail Feigin${}^*$,
Olaf Lechtenfeld${}^{\,\times}$ 
and \ Alexios P. Polychronakos${}^{\,\circ}$ }\\[8mm]

\noindent ${}^*${\em
School of Mathematics and Statistics, University of Glasgow\\
15 University Gardens, Glasgow G12 8QW, U.K. } \\
{Email: misha.feigin@glasgow.ac.uk}\\[6mm]

\noindent ${}^\times${\em
Institut f\"ur Theoretische Physik\/} \ and \ {\em
Riemann Center for Geometry and Physics\\
Leibniz Universit\"at Hannover \\
Appelstra\ss{}e 2, 30167 Hannover, Germany }\\
{Email: lechtenf@itp.uni-hannover.de}\\[6mm]

\noindent ${}^{\circ}${\em
CCPP, Department of Physics, NYU \\
4 Washington Pl., New York, NY 10016, USA \\
\vskip 0.1cm
{\rm{and}} \\
\vskip 0.1cm
Physics Department, The City College of the CUNY \\
160 Convent Avenue, New York, NY 10031, USA }\\
{Email: alexios@sci.ccny.cuny.edu}

\vspace{12mm}

\begin{abstract}
\noindent
The rational Calogero--Moser model of $n$ one-dimensional 
quantum particles with inverse-square pairwise interactions 
(in a confining harmonic potential) is reduced along the
radial coordinate of~$\R^n$ to the `angular Calogero--Moser
model' on the sphere~$S^{n-1}$. We discuss the energy
spectrum of this quantum system, its degeneracies 
and the eigenstates. The spectral flow with the coupling
parameter yields isospectrality for integer increments.
Decoupling the center of mass before effecting the spherical
reduction produces a `relative angular Calogero--Moser model',
which is analyzed in parallel. We generalize our considerations 
to the Calogero--Moser models associated with Coxeter groups.
Finally, we attach spin degrees of freedom to our particles 
and extend the results to the spin--Calogero system.
\end{abstract}
\end{center}

\end{titlepage}

\section{Introduction and summary}

\noindent
The Calogero model \cite{Cal} and its generalizations, called Calogero--Moser
models, constitute the paradigm of integrable 
and solvable multi-particle systems in one space dimension 
(for reviews, see~\cite{olpe1,olpe2,polyrev}). 
They have been applied to fluid mechanics, spin chains, gauge theory 
and string theory, and seemingly all their aspects have been 
thoroughly analyzed. 
Despite the long history of the subject~\cite{scholarpedia}, however, 
there still appear to be untrodden paths emanating.
One of these, a codimension-one reduction of the rational 
Calogero--Moser model to an integrable system on a sphere,
is the subject of this paper.

One may view $n$ interacting particles in one dimension
as a single particle in $\R^n$ moving in a specific potential 
(and possibly seeing a nontrivial metric).
Due to the conformal invariance of the rational Calogero--Moser model,
with the dilatation operator generating a rescaling of the 
radial coordinate~$r$ in~$\R^n$, a reduction of the model 
to $S^{n-1}$ is natural.\footnote{
Likewise, the translational symmetry of the model allows for 
a reduction to $\R^{n-1}$, which decouples the center of mass.
We also treat this reduction and combine the two.}
Indeed, upon passing to polar coordinates in~$\R^n$, 
all dependence on angular coordinates or momenta in the 
Hamiltonian~$H$ resides in a `deformed' centrifugal barrier term. 
Its $r^{-2}$ coefficient equals the quadratic Casimir of the conformal 
$sl(2)$ algebra (shifted by $\sfrac38$) and defines the Hamiltonian~$H_\Om$ 
of a reduced Calogero--Moser system living on the sphere~$S^{n-1}$,
which we call the (rational) {\it angular Calogero--Moser model\/}.
A variant of this reduction first separates the center of mass from
the particles' relative motion before applying the polar 
decomposition (then in $\R^{n-1}$), arriving at a {\it relative angular
Calogero--Moser model\/} defined on~$S^{n-2}$.

Various classical properties of these reduced models have already been 
investigated, such as their superintegrability~\cite{HNY,HKLN}, 
their conserved charges~\cite{HLNS,HLN} 
or their angle-action variable representation~\cite{LNY,HLNSY}. 
Intertwining relations for the quantum models were studied in \cite{Feig}.

In the current paper, we deal with the fundamental properties of the 
quantum models and focus on energy spectra, their degeneracies and
eigenstates. We render the energy spectra discrete
by adding a (one-body) harmonic confining potential, a harmless
modification which preserves integrability.
It turns out that the energy levels are easy to obtain, while
one has to work a bit harder for the eigenfunctions. We present two 
different ways to constructing them: firstly, by projecting the full model's
eigenstates onto its radial-oscillator ground state and, secondly,
by computing appropriate homogeneous symmetric polynomials with the help
of Dunkl operators. At integer values for the Calogero--Moser coupling 
constant magic happens: the energy levels are those of a free particle
on the sphere, but with degeneracies given by an effective angular momentum
which gets diminished with growing coupling. We also extend this analysis 
to Calogero--Moser models associated with Coxeter groups \cite{olpe2}.

The issue of degeneracies becomes richer when one attaches `spin' variables
to each particle. These so-called `spin--Calogero models'~\cite{spinCal} 
may equally be spherically reduced to obtain angular versions endowed with
spin degrees of freedom. We present the calculation of their degeneracies.

Part of our results (for the relative model) can be discovered already in 
the appendices of the original Calogero paper~\cite{Cal} 
(see also \cite{Cal2,Cal3} for $n{=}3$). 
We include them here for a more modern exposition and their relation 
to the mathematical framework developed later.

\bigskip

\section{The angular Calogero--Moser model}

\noindent
The prototypical integrable multi-particle system in one space dimension is the
Calogero--Moser model, whose two-body interaction comes in a rational, trigonometric,
hyperbolic or elliptic version. Here, we restrict ourselves to the rational model and
include a common (one-body) harmonic confining potential in order to obtain a
discrete energy spectrum in the quantum theory.

We parametrize the $n$-particle quantum phase space with coordinates $x^i$ and
momenta $p_j$, subject to the canonical commutation relations (setting $\hbar=1$)
\begin{equation}
[\,x^i\,,\,p_j\,] \= \im\,\delta^i_{\ j} \quad\with i,j=1,\ldots,n\ ,
\end{equation}
and introduce a radial coordinate~$r$ and momentum~$p_r$ via
\begin{equation}
\sum_{i=1}^n (x^i)^2 \= r^2 \und
\sum_{i=1}^n p_i^2 \= p_r^2 + \sfrac{1}{r^2}L^2 + \sfrac{(n-1)(n-3)}{4\,r^2}
\end{equation}
with $L^2=\sum_{i<j}(x^ip_j{-}x^jp_i)^2$ being the Laplacian on $S^{n-1}$.
The quantum Hamiltonian reads
\begin{equation} \label{qH}
\begin{aligned}
H &\= \sfrac12 \sum_i p_i^2\ +\ \sfrac{\o^2}{2}\sum_i (x^i)^2\ +\
\sum_{i<j} \frac{g(g{-}1)}{(x^i{-}x^j)^2} \\
&\= \sfrac12 p_r^2\ +\ \sfrac{\o^2}{2}r^2\ +\ \sfrac{(n-1)(n-3)}{8\,r^2}\ +\
\sfrac{1}{r^2} H_\Om \\[8pt]
&\= \sfrac{1}{2n} P^2\ +\ \sfrac{n\o^2}{2} X^2\ +\ \tiH \\[8pt]
&\= \sfrac{1}{2n} P^2\ +\ \sfrac{n\o^2}{2} X^2\ +\
\sfrac12\tip_r^2\ +\ \sfrac{\o^2}{2}\tir^2\ +\ \sfrac{(n-2)(n-4)}{8\,\tir^2}\ +\
\sfrac{1}{\tir^2}\tiH_\Om\ ,
\end{aligned}
\end{equation}
where we introduced the center-of-mass coordinate and momentum
\begin{equation}
X\=\sfrac1n\sum_i x^i \und P\=\sum_i p_i\ ,
\end{equation}
respectively, as well as a relative Hamiltonian~$\tiH$, 
radial coordinate~$\tir$ and momentum~$\tip_r$ via
\begin{equation}
\sfrac1n\sum_{i<j} (x^i{-}x^j)^2 \= \tir^2 \und
\sfrac1n\sum_{i<j} (p_i{-}p_j)^2 \= 
\tip_r^2 + \sfrac{1}{\tir^2}\tiL^2 + \sfrac{(n-2)(n-4)}{4\,\tir^2}\ ,
\end{equation}
with $\tip_r$ canonically conjugate to $\tir$ and $\tiL^2$ defined analogously to~$L^2$.
A useful relation is \ $r^2=n\,X^2+\tir^2$.

The implicitly given operators $H_\Om$ and $\tiH_\Om$ are 
named `angular Calogero--Moser Hamiltonian'
and `relative angular Calogero--Moser Hamiltonian', 
since they define integrable submodels independent of the
radial and relative radial degree of freedom, 
living on the spheres $S^{n-1}$ and $S^{n-2}$, respectively.
Their classical properties have been thoroughly investigated 
in~\cite{HNY,HKLN,HLNS,HLN,LNY,HLNSY}.
In the present paper, we focus on their quantum features, 
in particular their energy spectrum and eigenstates.
Explicitly, these Hamiltonians read
\begin{eqnarray} \label{qHom}
H_\Om &=& \sfrac12 L^2 \ +\ r^2 \sum_{i<j} \frac{g(g{-}1)}{(x^i{-}x^j)^2} 
\ \=\ \sfrac12 L^2 \ +\ \sfrac{g(g{-}1)}{2} \Bigl( \sfrac{n(n{-}1)}{2} +
\sum_{\alpha\in{\cal R}_+} \tan^2\th_\alpha \Bigr)\ ,\\[8pt]
\tiH_\Om &=& \sfrac12 \tiL^2 \ +\ \tir^2 \sum_{i<j} \frac{g(g{-}1)}{(x^i{-}x^j)^2} 
\ \=\ \sfrac12 \tiL^2 \ +\ \sfrac{g(g{-}1)}{2} \Bigl( \sfrac{n(n{-}1)}{2} +
\sum_{\alpha\in{\cal R}_+} \tan^2\widetilde\th_\alpha \Bigr)\ ,
\end{eqnarray}
where the sum runs over the positive roots of $A_{n-1}$, 
and $\th_\alpha$ ($\widetilde\th_\alpha$)
is the angle of the unit vector $\th\in\R^n$ ($\widetilde\th\in\R^{n-1}$)
with the root~$\alpha$.
Because $L^2$ ($\tiL^2$) is the Laplacian on the sphere $S^{n-1}$ ($S^{n-2}$),
these Hamiltonians describe particle motion on that sphere in the presence of specific
potentials, which become singular on the intersections with the Weyl chamber walls.
These potentials are sometimes called `Higgs oscillators'~\cite{higgs,leemon}.
Note that the frequency~$\o$ does not appear here. The confining potential only
serves to render the full Calogero--Moser spectrum discrete and may be turned off 
when convenient. We will assume $g\ge0$ without loss of generality.

On position wave functions, the momentum operators~$p_i$ are represented by $-\im\pa_i$, 
hence
\begin{equation}
P\ \mapsto\ -\im\pa_X \und
p_r\ \mapsto\ -\im\bigl(\pa_r + \sfrac{n-1}{2\,r}\bigr) \quad,\qquad
\tip_r\ \mapsto\ -\im\bigl(\pa_{\tir} + \sfrac{n-2}{2\,\tir}\bigr) \ ,
\end{equation}
and therefore
\begin{equation}
\begin{aligned}
H &\ \mapsto\ -\sfrac12 \pa_r^2\ -\ \sfrac{n-1}{2\,r}\pa_r\ +\ \sfrac{\o^2}{2}r^2\ +\
\sfrac{1}{r^2} H_\Om \\[8pt]
&\= -\sfrac{1}{2n} \pa_X^2\ +\ \sfrac{n\o^2}{2} X^2\ -\
\sfrac12\pa_{\tir}^2\ -\ \sfrac{n-2}{2\,\tir}\pa_{\tir}\ +\ \sfrac{\o^2}{2}\tir^2\ +\
\sfrac{1}{\tir^2}\tiH_\Om\ ,
\end{aligned}
\end{equation}
Hence, on these wave functions, we can represent
\begin{equation} \label{Homrep}
\begin{aligned}
H_\Om &\= r^2 H - \sfrac{\o^2}{2}r^4\ +\ \sfrac12(r\pa_r+n{-}2)\,r\pa_r \ ,\\[4pt]
\tiH_\Om &\=
\tir^2\tiH - \sfrac{\o^2}{2}\tir^4\ +\ \sfrac12(\tir\pa_{\tir}+n{-}3)\,\tir\pa_{\tir} \ .
\end{aligned}
\end{equation}

The standard similarity transformation
(with $\th\in S^{d-1}$ and $\widetilde\th\in S^{d-2}$)
\begin{equation}
\Psi(r,\th) \= r^{-\frac{n-1}{2}} \, u(r,\th) \qquad{\rm or}\qquad
\tipsi(X,\tir,\widetilde\th) \= \tir^{-\frac{n-2}{2}} \,\tiu(\tir,\widetilde\th)\,\chi(X)
\end{equation}
removes the first-derivative term at the expense of a constant shift,
\begin{equation}
\begin{aligned}
H &\ \mapsto\ -\sfrac12 \pa_r^2\ +\ \sfrac{\o^2}{2}r^2\ +\ \sfrac{(n-1)(n-3)}{8\,r^2}\ +\
\sfrac{1}{r^2} H_\Om \\[8pt]
&\= -\sfrac{1}{2n} \pa_X^2\ +\ \sfrac{n\o^2}{2} X^2\ -\
\sfrac12\pa_{\tir}^2\ +\ \sfrac{\o^2}{2}\tir^2\ +\ \sfrac{(n-2)(n-4)}{8\,\tir^2}\ +\
\sfrac{1}{\tir^2}\tiH_\Om\ ,
\end{aligned}
\end{equation}
which amounts simply to
\begin{equation}
P\ \mapsto\ -\im\pa_X \und
p_r\ \mapsto\ -\im\,\pa_r \quad,\qquad
\tip_r\ \mapsto\ -\im\,\pa_{\tir}
\end{equation}
in~(\ref{qH}).

\bigskip

\section{Energy spectra}

\noindent
In order to find the spectra of $H_\Om$ and $\tiH_\Om$, we define the following
eigenvalue problems,
\begin{eqnarray}
H\,u_k \= E_k\,u_k \quad &,& \qquad \tiH\,\tiu_k \= \tiE_k\,\tiu_k\ , \\
H_\Om\,v_k \= \ve_k\,v_k \quad &,& \qquad \tiH_\Om\,\tiv_k \= \tie_k\,\tiv_k\ ,
\end{eqnarray}
where $k$ collectively counts the discrete eigenvalues.

The standard Calogero--Moser spectrum in the presence of a harmonic term is a classic
result~\cite{Cal,olpe2,polyrev},
\begin{equation} \label{E}
E_k \= \o\,\bigl(\sfrac12 g\,n(n{-}1)+\sfrac{n}{2}+k_1+2k_2+3k_3+\ldots+nk_n\bigr)
\quad\with k_i=0,1,2,\ldots\ .
\end{equation}
Since the center-of-mass Hamiltonian is a standard harmonic oscillator,
\begin{equation}
\bigl( -\sfrac{1}{2n} \pa_X^2 + \sfrac{n\o^2}{2} X^2 \bigr)\,\chi_k(X) \=
\o\,(k{+}\sfrac12)\,\chi_k(X)\ ,
\end{equation}
its contribution in~(\ref{E}) is identified with the $k_1$ term (and a constant~$\sfrac12\o$).
Hence, the relative Calogero--Moser spectrum must just lack the $k_1$ term,
\begin{equation} \label{tiE}
\tiE_k \= \o\,\bigl(\sfrac12 g\,n(n{-}1)+\sfrac{n-1}{2}+2k_2+3k_3+\ldots+nk_n\bigr)
\quad\with k_i=0,1,2,\ldots\ .
\end{equation}

A simple trick reveals the spectra of $H_\Om$ and $\tiH_\Om$.
According to~(\ref{qH}), on any $H_\Om$ eigenspace for eigenvalue~$\ve_k$,
the full Calogero--Moser Hamiltonian reduces to
\begin{equation}
H\big|_{\ve_k} \= \sfrac12 p_r^2\ +\ \sfrac{\o^2}{2}r^2\ +\ \sfrac{(n-1)(n-3)}{8\,r^2}\ +\
\sfrac{\ve_k}{r^2} \= \sfrac12 p_r^2 + \sfrac{\o^2}{2}r^2 + \sfrac{h(h-1)}{2\,r^2}\ ,
\end{equation}
reparametrizing
\begin{equation}\label{hdef}
\ve_k + \sfrac18(n{-}1)(n{-}3) \ =:\ \sfrac12 h(h{-}1)\ .
\end{equation}
However, this is just the relative two-particle Calogero--Moser Hamiltonian~$\tiH$ 
for $n{=}2$, as can be seen from (\ref{qH}) and~(\ref{qHom}) with
\begin{equation}
\tiL^2=0 \quad,\qquad \sfrac12(x^1{-}x^2)^2 = \tir^2 \quad,\qquad g\ \to\ h
\end{equation}
and the tildes removed. Thus, we know from~(\ref{tiE}) that
\begin{equation}
{\rm spec}\,\bigl(H\big|_{\ve_k}\bigr) \=
\bigl\{ \o\,(h+\sfrac12+2k_2) \with k_2=0,1,2,\ldots \bigr\} \ ,
\end{equation}
and comparing to~(\ref{E}) yields the identification
\begin{equation}
h \= \sfrac12 g\,n(n{-}1)+\sfrac{n-1}{2}+k_1+3k_3+4k_4+\ldots+nk_n
\ =:\ q+\sfrac{n-1}{2}\ .
\end{equation}
Note that the $k_2$ term has disappeared from~$h$. From~(\ref{hdef}) we now read off that
\begin{equation}\label{ve}
\ve_k\=\sfrac12 h(h{-}1)-\sfrac18(n{-}1)(n{-}3) \= \sfrac12 q\,(q+n-2)
\end{equation}
with
\begin{equation}\label{q}
q\=\sfrac12 g\,n(n{-}1)+k_1+3k_3+4k_4+\ldots+nk_n\ .
\end{equation}
We see that, while the full Calogero--Moser energy~$E_k$ depends linearly 
on the $n$ quantum numbers $k_1,k_2,k_3\ldots,k_n$, the angular Calogero--Moser 
energy~$\ve_k$ is independent of~$k_2$ and depends quadratically on the 
remaining~$k_i$ via the combination~$q$.
For integer values of the coupling~$g$, the angular energy spectrum~(\ref{ve}) is that
of a free particle with angular momentum~$q$ on the $(n{-}1)$-sphere, but has a lower
degeneracy as we shall see. Each time $g$ is increased by one, $q$ grows by $\frac12n(n{-}1)$,
so the lowest $\frac12n(n{-}1)$ energy levels disappear from the spectrum while the rest is
reproduced, with a reduced degeneracy.

This story gets repeated for the relative angular Calogero--Moser energy. We have
\begin{equation} \label{tihdef}
\tiH\big|_{\tie_k} \= \sfrac12\tip_r^2\ +\ \sfrac{\o^2}{2}\tir^2\ +\ 
\sfrac{\tih(\tih-1)}{2\,\tir^2}
\quad\with \sfrac12\tih(\tih{-}1)\=\tie_k+\sfrac18(n{-}2)(n{-}4)\ ,
\end{equation}
and, hence, obtain from~(\ref{tiE}) the identification
\begin{equation}
\tih \= \sfrac12 g\,n(n{-}1)+\sfrac{n-2}{2}+3k_3+4k_4+\ldots+nk_n
\ =:\ \tiq+\sfrac{n-2}{2}\ .
\end{equation}
Both $k_1$ and $k_2$ are absent from~$\tih$. From~(\ref{tihdef}) we finally extract
\begin{equation} \label{tie}
\tie_k\=\sfrac12 \tih(\tih{-}1)-\sfrac18(n{-}2)(n{-}4) \= \sfrac12\tiq\,(\tiq+n-3)
\end{equation}
where
\begin{equation} \label{tiq}
\tiq\=\sfrac12 g\,n(n{-}1)+3k_3+4k_4+\ldots+nk_n \= q-k_1\ .
\end{equation}
The relative angular Calogero--Moser energy depends quadratically only on the 
quantum numbers $k_3,k_4,\ldots,k_n$ of the full Calogero--Moser model.
For comparison, we rewrite the full absolute and relative Calogero--Moser energies,
\begin{equation}
E_k\= \o\bigl(q+\sfrac{n}{2}+2k_2\bigr) \und
\tiE_k\= \o\bigl(\tiq+\sfrac{n-1}{2}+2k_2\bigr) \ ,
\end{equation}
in terms of $q$ and $\tiq$, respectively.

To summarize, the center of mass and the (absolute or relative) radial degree of freedom
in the Calogero--Moser model account for the $k_1$ and $k_2$ quantum numbers of the 
energy eigenstates. Setting them to zero provides the quantum numbers for the relative 
angular Calogero--Moser model, whose energy depends quadratically on the 
combination~(\ref{tiq}) in~(\ref{tie}).

It is known that the energy eigenstates are uniquely characterized by the multiindex~$k$.
Let us define its level number~$m(k)$ by
\begin{equation}
m\=k_1+2k_2+3k_3+4k_4+\ldots+nk_n \quad\for
k=(k_1,k_2,k_3,k_4,\ldots,k_n) \ .
\end{equation}
Since only one branch of the quadratic functions (\ref{ve}) and~(\ref{tie}) is relevant,
the degeneracy of a given energy eigenvalue is given by the number $p_n(m)$ of partitions
of~$m$ into integers not bigger than~$n$. The only difference between the four models is that,
in the relative cases, the `ones' are excluded from the partitions, while for the angular
models the `twos' have to be omitted. Therefore, the respective degeneracies are given by
\begin{equation} \label{alldeg}
p_n(m)\ ,\quad p_n(m)-p_n(m{-}1)\ ,\quad
p_n(m)-p_n(m{-}2)\ ,\quad p_n(m)-p_n(m{-}1)-p_n(m{-}2)+p_n(m{-}3)\ .
\end{equation}

We close this section by giving explicit formulae 
for the special cases of $n{=}2$ and $n{=}3$.
In the two-particle system, our equations degenerate to $q=g{+}k_1$ and $\tiq=g$, thus
\begin{equation}
E_k=\o(g{+}1{+}k_1{+}2k_2)\ ,\quad \tiE_k=\o(g{+}\sfrac12{+}2k_2)\ ,\quad
\ve_k=\sfrac12 q^2\ ,\quad \tie_k=\sfrac12 \tiq(\tiq{-}1)\ .
\end{equation}
As the relative angular two-body Calogero--Moser system is empty, 
its energy reduces to a constant.
For three particles, one has $q=3g{+}k_1{+}3k_3$ and $\tiq=3g{+}3k_3$, and therefore
\begin{equation}
E_k=\o(3g+\sfrac32+k_1+2k_2+3k_3)\ ,\quad \tiE_k=\o(3g+1+2k_2+3k_3)\ ,\quad
\ve_k=\sfrac12 q(q{+}1) \ ,\quad \tie_k=\sfrac12\tiq^2\ ,
\end{equation}
i.e.~$\tie_k=\sfrac92(g{+}k_3)^2$, which is known from the P\"oschl-Teller model.
No simplifications appear to occur for~$n\ge4$.

\bigskip

\section{Energy eigenstates}

\noindent
It is well known how to construct the energy eigenfunctions~\footnote{
Arguments $x$ and $p$ stand for $\{x^1,x^2,\ldots,x^n\}$ and $\{p_1,p_2,\ldots,p_n\}$,
respectively. Similarly for $\tix$ and $\tip$.} \
$\Psi_k(x)=\<x\,|\,k_1,k_2,k_3,\ldots,k_n\>$ \ of the rational Calogero--Moser model
with common harmonic potential, for any energy eigenstate $|k\>=|k_1,k_2,k_3,\ldots,k_n\>$.

The ground-state wave function reads~\footnote{
Here and below, we do not normalize states or wave functions.
$k=0$ means $k=(0,0,\ldots,0)$.}
\begin{equation}
\Psi_{0}\= \Delta^{g}\,\ep^{-\frac12\o r^2}
\= \Delta^g\,\ep^{-\frac12\o\tir^2}\ep^{-\frac{n}{2}\o X^2}
\= \tipsi_0(\tix)\,\chi_0(X)
\quad\with\quad \Delta\=\prod_{i<j}(x^i{-}x^j)
\end{equation}
and is annihilated by
\begin{equation}
A_\ell\=I_\ell(p{-}\im\o x,x) \for \ell=1,2,3,\ldots,n\ ,
\end{equation}
where $I_\ell(p,x)$ is the $\ell$th-order conserved charge 
of the scattering Calogero--Moser model (without harmonic term),
given by the matrix trace of the $\ell$th power 
of the $n{\times}n$ matrix-valued Lax operator~\cite{olpe2}.
The creation operators
\begin{equation}
A_\ell^\+\=I_\ell(p{+}\im\o x,x) \for \ell=1,2,3,\ldots,n
\end{equation}
commute with one another and help build all excited states,
\begin{equation} \label{basis}
|k_1,k_2,k_3,\ldots,k_n\> \=
\bigl(A_1^\+\bigr)^{k_1}\bigl(A_2^\+\bigr)^{k_2}\cdots\bigl(A_n^\+\bigr)^{k_n}\,
|0,0,\ldots,0\>\ .
\end{equation}
Employing
\begin{equation}
(-\pa_i{+}\o x^i)\,\Psi_0\=\bigl(2\o x^i-g\sum_{j(\neq i)}(x^i{-}x^j)^{-1}\bigr)\,\Psi_0
\end{equation}
and some identities on partial fractions it is straightforward 
but cumbersome to compute any~$\Psi_k$.

For illustration, the first three conserved charges take the form
\begin{equation}
\begin{aligned}
I_1(p,x)&\=\sum_i p_i\=P\ ,\qquad
I_2(p,x)  \=\sum_i p_i^2 + 2g(g{-}1)\sum_{i<j}(x^i{-}x^j)^{-2} \=2H\ ,\\
I_3(p,x)&\=\sum_i p_i^3 + 3g(g{-}1)\sum_{i<j}(x^i{-}x^j)^{-2}(p_i{+}p_j)\ ,
\end{aligned}
\end{equation}
and, therefore,
\begin{equation}
\begin{aligned}
A_1^\+&\=P+\im\o nX\ ,\qquad
A_2^\+  \=2H+\im\o(x{\cdot}p+p{\cdot}x)-2\o^2 r^2\ ,\\[4pt]
A_3^\+&\=\sum_i (p_i{+}i\o x^i)^3 +
3g(g{-}1)\sum_{i<j}(x^i{-}x^j)^{-2}(p_i{+}p_j+\im\o x^i{+}\im\o x^j)\ ,
\end{aligned}
\end{equation}
leading to
\begin{eqnarray}
\Psi_{100\ldots0}&=& 2\im\o n X\,\Psi_0 \ ,\\
\Psi_{200\ldots0}&=& 2\o n\,(1-2\o nX^2)\,\Psi_0\ ,\\
\Psi_{300\ldots0}&=& 4\im\o^2 n^2 (3X-2\o nX^3)\,\Psi_0 \ ,\\
\Psi_{010\ldots0}&=& \bigl( 2\o gn(n{-}1)+2\o n- 4\o^2 r^2\bigr)\,\Psi_0\ ,\\
\Psi_{110\ldots0}&=& 4\im\o^2 nX\,\bigl(gn(n{-}1)+(n{+}2)-2\o r^2\bigr)\,\Psi_0\ ,\\
\Psi_{001\ldots0}&=& \bigl(12\im\o^2n[1{+}g(n{-}1)]X - 
8\im\o^3{\textstyle\sum_i}(x^i)^3\bigr)\,\Psi_0\ .
\end{eqnarray}

An equivalent strategy makes use of the exchange-creation operators \cite{exop,BHV}
\begin{equation} \label{excha}
a_i^\dagger \= -\im \cD_i + \im \o x^i \und a_i \= -\im \cD_i -\im \o x^i\ ,
\end{equation}
constructed using Dunkl operators
\begin{equation} \label{dunkl}
\cD_i \= \pa_i\ +\ g\sum_{j(\neq i)}\frac{1-s_{ij}}{x^i-x^j}\ ,
\end{equation}
which contain permutation operators~$s_{ij}$, commute with each other
and map polynomials to polynomials.
Employing the totally symmetric Newton sums
\begin{equation}
B_\ell^\+\= \sum_i ( a_i^\dagger )^\ell\ ,
\end{equation}
we may alternatively write
\begin{equation}
\Psi_k(x) \= \Delta^g
\bigl(B_1^\+\bigr)^{k_1}\bigl(B_2^\+\bigr)^{k_2}\cdots\bigl(B_n^\+\bigr)^{k_n}\,
\ep^{-\frac12\o r^2}\ .
\end{equation}
Each wave function is a totally symmetric polynomial of degree
$m=\sum_i ik_i$ multiplying~$\Psi_0$.
Due to the invariance under permutations of the coordinates~$x^i$,
the degeneracy at a given level~$m$ is much smaller than that of an $n$-dimensional
isotropic harmonic oscillator.

How can one extract from these full eigenfunctions the eigenfunctions \
$\tipsi_k=\tir^{-\frac{n-2}2}\tiu_k$, \\ $v_k$ and $\tiv_k$ of the other Hamiltonians?
This task is rather straightforward for $\tipsi_k$:
For any given multiindex
\begin{equation} \label{knok1}
k=(0,k_2,k_3,\ldots,k_n) \quad\with 2k_2+3k_3+\ldots+nk_n=m\ ,
\end{equation}
rewrite the corresponding $\Psi_k$ in terms of $X$ and $\tix$,
then expand its $X$~dependence into the
harmonic-oscillator basis $\{\chi_{k'_1}\}$ of the center of mass,
\begin{equation}
\Psi_k(X,\tix) \=
\sum_{k'} c_k^{k'}\,\chi_{k'_1}(X)\,\tipsi_{0\,k'_2k'_3\ldots k'_n}(\tix)
\quad\with \sum_{i=1}^n ik'_i=m\ .
\end{equation}
In each term, the energy is split as
\begin{equation}
E \= \o\,\bigl( gn(n{-}1)+\sfrac{n}{2}+m\bigr) \=
\o\,\bigl( \sfrac12 + k'_1 \bigr)\ +\
\o\,\bigl( gn(n{-}1)+\sfrac{n-1}{2}+m-k'_1\bigr)\ .
\end{equation}
The coefficient of $\chi_0$ may be identified with $\tipsi_k$;
it lives in the relative eigenspace with $\tiE=E{-}\frac{\o}{2}$.
This relative eigenspace is spanned by all $\tipsi_{0,k_2,k_3,\ldots,k_n}$
obtained from partitions~(\ref{knok1}).
In other words, we select those Calogero--Moser states whose center-of-mass
oscillator is not excited.
The simplest example is $m{=}2$,
\begin{equation}
\begin{aligned}
\Psi_{010\ldots0} &\= \bigl( 2\o gn(n{-}1)+2\o n- 4\o^2 r^2\bigr)\,\Psi_0 \\[2pt]
&\= \bigl( 2\o gn(n{-}1)+2\o(n{-}1)-4\o^2\tir^2\bigr)\,\Psi_0\ +\
\bigl(2\o-4\o^2 nX^2\bigr)\,\Psi_0\\
&\= \chi_0\,\tipsi_{010\ldots0}\ +\ \chi_2\,\tipsi_{000\ldots0}
\end{aligned}
\end{equation}
with \ $\chi_2=2\o(1{-}2\o nX^2)\chi_0$, thus
\begin{equation}
\tipsi_{010\ldots0}\ \propto\ \bigl( (gn{+}1)(n{-}1)-2\o\tir^2\bigr)\,\tipsi_0\ .
\end{equation}

The situation is more involved for the eigenfunctions~$v_k$, where now
\begin{equation} \label{vk}
k=(k_1,0,k_3,k_4,\ldots,k_n) \quad\with k_1+3k_3+4k_4+\ldots+nk_n=m\ .
\end{equation}
To derive these at a given energy level $E=\o(gn(n{-}1)+\sfrac{n}{2}+m)$,
we have to rewrite
\begin{equation}
\Psi_k(x) \=  r^{-\frac{n-1}{2}} u_k(r,\th)
\end{equation}
in terms of the radial~($r$) and angular~($\th\in S^{d-1}$) coordinates.
The $r$~dependence of $u_k$ has to be expanded in terms
of the solutions of the following eigenvalue problem,
\begin{equation}
\bigl(-\sfrac12\pa_r^2+\sfrac12\o^2 r^2+\sfrac{h(h-1)}{2\,r^2}\bigr)\,\rho_{k'_2}(r)
\=E_{k'_2}\,\rho_{k'_2}(r)
\quad\with E_{k'_2}=\o\bigl(h+\sfrac12+2k'_2\bigr) \ .
\end{equation}
This is nothing but the (reduced) radial wave equation for the isotropic harmonic 
oscillator in $n$~dimensions, with angular momentum~$q=h-\sfrac{n-1}{2}$ and
principal quantum number~$k'_2$. Hence, we know that
\begin{equation}
\rho_{k'_2}(r)\ \propto\ r^h\,\ep^{-\frac12\o r^2} F(-k'_2,h{+}\sfrac12;\o r^2)
\ \propto\ r^h\,\ep^{-\frac12\o r^2} L_{k'_2}^{h-\frac12}(\o r^2)\ ,
\end{equation}
involving the confluent hypergeometric function~$F$
or associated Laguerre polynomials~$L_n^\a$,
\begin{equation}
F(-n,\b;z) \= \sum_{j=0}^n \frac{(-1)^j\,n!\,\Gamma(\b)}
{(n{-}j)!\,j!\,\Gamma(\b{+}j)} \,z^j \qquad{\rm or}\qquad
L_n^\a(z) \= \sfrac1{n!}\,z^{-\a}\ep^z\,
\frac{\diff^n}{\diff z^n}\bigl(\ep^{-z}z^{n+\a}\bigr)\ .
\end{equation}

In each term of the expansion
\begin{equation} \label{uexp}
u_k(r,\th)\= \sum_{k'} d_k^{k'}\,\rho_{k'_2}(r)\,v_{k'_1\,0\,k'_3k'_4\ldots k'_n}(\th)
\quad\with \sum_{i=1}^n ik'_i=m\ ,
\end{equation}
the value of $h$ belonging to~$\rho_{k'_2}$ is fixed by the requirement
\begin{equation}
E_{k'_2}=E_k \qquad\Longrightarrow\qquad
h\=\sfrac12 gn(n{-}1)+\sfrac{n-1}{2}+m-2k'_2\ .
\end{equation}
The coefficient of $\rho_0\!=\!r^h\ep^{-\frac12\o r^2}$
may be identified with $v_k(\th)$,which amounts to setting
$d_k^{k'}=\delta_{kk'}$ for $k'_2{=}0$.
In other words, $v_k$ is the projection of $u_k$ to its radial oscillator
ground state (with the proper value of~$h$).
As a check, $v_k$ should not depend nontrivially on~$\o$.
By collecting the $v_k$ pertaining to all partitions of $m$ in~(\ref{vk}),
we span the eigenspace of~$H_\Om$ to the energy
\begin{equation} \label{epsq}
\ve\=\sfrac12q\,(q+n-2) \quad\with q=\sfrac12g\,n(n{-}1)+m\ .
\end{equation}

Let us consider the first nontrivial example,
\begin{equation}
k=(2,0,0,\ldots,0) \qquad\Longrightarrow\qquad
m=2 \and k'_2=0,1\ .
\end{equation}
The task is to expand
\begin{equation}
u_{20\ldots0}\=2\o n\,(1-2\o nX^2)\,\D^g\,r^{\frac{n-1}{2}}\,\ep^{-\frac12\o r^2}
\end{equation}
into
\begin{equation}
\rho_0=r^2\rho \and
\rho_1=\bigl((\sfrac12gn(n{-}1){+}\sfrac{n}{2})-\o r^2\bigr)\rho
\qquad{\rm with}\quad
\rho=r^{\frac12gn(n-1)+\frac{n-1}2}\,\ep^{-\frac12\o r^2}
\end{equation}
as
\begin{equation}
u_{20\ldots0}(r,\th)\=\rho_0(r)\,v_{20\ldots0}(\th)\ +\ \rho_1(r)\,v_{00\ldots0}(\th)\ ,
\end{equation}
where we are allowed to absorb expansion coefficients into the~$v_{k'}$.
After splitting
\begin{equation}
X\=r\,\widehat{X} \und \Delta\=r^{\frac12n(n-1)}\,\widehat{\Delta}
\end{equation}
and matching powers of $r$, we read off that
\begin{equation}
2\o n\,\widehat{\D}^g \= \bigl(\sfrac12gn(n{-}1){+}\sfrac{n}2\bigr)\,v_{00\ldots0}
\und -4\o^2 n^2 \widehat{X}^2 \widehat{\D}^g \= v_{20\ldots0} - \o\,v_{00\ldots0}
\end{equation}
and arrive at
\begin{equation} \label{vkex}
v_{00\ldots0}\ \propto\ \widehat{\D}^g \und
v_{20\ldots0}\ \propto\ \bigl\{1-[g(n{-}1){+}1]\,n^2\widehat{X}^2 \bigr\}\,\widehat{\D}^g\ .
\end{equation}
Since only the first two modes are excited, the angular dependence above the ground state
is expressed through~$\widehat{X}$ alone.
Hence, we may rotate our coordinate system such that only a single angle~$\g$ appears,
$n\widehat{X}^2=\cos^2\g$. Already for $m{=}3$, however, $k_3$ will be turned on,
and the complexity increases noticeably.

A more elegant way to extract the angular wave functions makes use of
the Dunkl operators~(\ref{dunkl}).
Since we do not need the normalizability of the radial wave functions,
we put $\o=0$ in the following.
To each angular eigenfunction $v_k(\th)$, we may associate a homogeneous polynomial
$h_k(x)$ of degree~$m$ via
\begin{equation} \label{hkdef}
v_k(\th) \= r^{-m} h_k(x)\ \widehat{\D}^g
\qquad\Longrightarrow\qquad
\Delta^g\,h_k(x) \= r^{\frac12gn(n{-}1)+m}v_k(\th) \= r^q v_k(\th)\ .
\end{equation}
Comparing
\begin{equation}
\sfrac12(r\pa_r+n{-}2)\,r\pa_r\,(r^q v_k) \= \sfrac12 q(q{+}n{-}2)\,r^q v_k \=
\ve_k\,r^q v_k \= H_\Om\,(r^q v_k)
\end{equation}
to the first line of~(\ref{Homrep}) at $\o{=}0$, we learn that
\begin{equation}
H\,(r^q v_k) \= H\,(\Delta^g\,h_k) \= 0\ .
\end{equation}
Introducing the Calogero--Moser operator  
\begin{equation}
L(g) \= \sum_i \pa_i^2  +
\sum_{i<j} \frac{2g}{x^i{-}x^j}(\partial_i{-}\partial_j)\ ,
\end{equation}
which on symmetric functions agrees with $\sum_i\cD_i^2$,
and employing the fact that
\begin{equation}
H\,\Delta^g \= -\half \Delta^g\,L(g)\ ,
\end{equation}
it follows that $h_k(x)$ must be a `deformed' harmonic polynomial,
\begin{equation} 
L(g)\,h_k(x) \= 0\ .
\end{equation}
Fortunately, the construction of such homogeneous degree~$m$ polynomials is 
known~\cite{dunkl}:\footnote{
For another method, see Section~3 of~\cite{vandiejen}.}
\begin{equation}
\begin{aligned} \label{defharm}
h_k(x) &\= r^{gn(n-1)+n-2+2m}\,(B_1^\+)^{k_1} (B_3^\+)^{k_3} \cdots (B_n^\+)^{k_n}\,
r^{-gn(n-1)-n+2} \\[4pt]
&\= r^{gn(n-1)+n-2+2m}\,({\textstyle\sum_i}\cD_i)^{k_1} ({\textstyle\sum_i}\cD_i^3)^{k_3}
\cdots ({\textstyle\sum_i}\cD_i^n)^{k_n}\,r^{-gn(n-1)-n+2}
\end{aligned}
\end{equation}
generates all of them.
It is easy to see that they are linearly independent.
Note that, again, $k_2{=}0$, because the Dunkl operators commute and
\begin{equation}
\sum_i \cD_i^2\,r^{-gn(n-1)-n+2} \= 0\ ,
\end{equation}
as is easily checked.

For instance, direct computation at $m{=}2$ produces
\begin{equation}
\begin{aligned}
h_{20\ldots0} &\= r^{gn(n-1)+n+2}\,({\textstyle\sum_i}\cD_i)^2\,r^{-gn(n-1)-n+2} \\
&\= r^{gn(n-1)+n+2}\,({\textstyle\sum_i}\pa_i)^2\,r^{-gn(n-1)-n+2}
\ \propto\ r^2-[g(n{-}1){+}1]\,n^2 X^2\ ,
\end{aligned}
\end{equation}
which, using~(\ref{hkdef}), yields $v_{20\ldots0}$ agreeing with~(\ref{vkex}).
As a less trivial example at $m{=}3$, we also present
\begin{equation}
\begin{aligned}
h_{0010\ldots0} &\= r^{gn(n-1)+n+4}\,{\textstyle\sum_i}\cD_i^3\,r^{-gn(n-1)-n+2} \\
&\ \propto\ 3[g(n{-}1){+}1]r^2 nX-[gn(n{-}1){+}n{+}2] {\textstyle\sum_i}(x^i)^3
\end{aligned}
\end{equation}
which yields
\begin{equation}
v_{0010\ldots0} \ \propto\ \bigl\{
3[g(n{-}1){+}1]\,n\widehat{X}-[gn(n{-}1){+}n{+}2]\,
{\textstyle\sum_i}(\widehat{x}^i)^3\bigr\}\,\widehat{\D}^g\ .
\end{equation}

The degeneracy of the energy $\ve(q)=\half q(q{+}n{-}2)$ with $q=\sfrac12gn(n{-}1)+m$
is given by the dimension of the space of certain harmonic homogeneous polynomials
of degree~$m$. If all such polynomials are considered, they furnish the totally
symmetric rank~$m$ representation of~SO$(n)$, whose dimension is given by
\begin{equation}
d(m) \= \bigl( \begin{smallmatrix} n+m-1 \\ n-1 \end{smallmatrix} \bigr)
\ -\ \bigl( \begin{smallmatrix} n+m-3 \\ n-1 \end{smallmatrix} \bigr)\ ,
\end{equation}
where the subtracted term accounts for removing the trace parts.
In our case, we are restricted to the subset of harmonic polynomials invariant
under the action of the permutation group~$S_n$, which reduces the dimension to
\begin{equation} \label{degsym}
d_{\rm sym}(m) \= p_n(m) - p_n(m{-}2)\ ,
\end{equation}
as stated earlier. In the free theory $(g{=}0)$, this is obvious since $m=q$.
Surprising looks the fact that, for any integer coupling~$g$, the degeneracy of the
energy~$\ve(q)$ is given by the same formula but with a diminished `effective
angular momentum' $m=q-\sfrac12gn(n{-}1)$.

This observation can also be explained with the help of intertwining operators~$K(g)$.
These intertwine two Calogero--Moser operators with couplings 
$g$ and $g{+}1$~\cite{Heckman}:
\begin{equation}\label{shift}
K(g)\,L(g{+}1) \= L(g)\,K(g)\ ,
\end{equation}
where $K(g)$ is a differential operator which has the form
\begin{equation}
K(g)\=\Bigl[\prod_{i<j} (\mathcal D_i-\mathcal D_j)\Bigr] \ \Delta\ 
\end{equation}
when acting on permutation-symmetric polynomials.
We note that the kernel of $K(g)$ does not contain such  polynomials. Indeed, 
let us suppose that there exists  a permutation-symmetric polynomial~$f$ with $K(g)f=0$. 
Then, for the inner product
\begin{equation}
(h',h) \= h'(\mathcal D_1,\ldots, \mathcal D_n)\,h(x^1,\ldots,x^n)\big|_{x=0}\ ,
\end{equation}
we have that
\begin{equation}
0 \= \bigl( K(g)f\,,\,h\bigr) \=
\Bigl( \Bigl[\prod_{i<j} (\mathcal D_i{-}\mathcal D_j)\Bigr] \,\Delta\,f\,,\,h\Bigr)
\= \big( \Delta\,f\,,\,\Delta\,h\bigr)
\end{equation}
because the adjoint of $\mathcal D_i$ is $\mathcal D_i^*=x^i$~\cite{DJO}.
Restricting $h$ to permutation-symmetric polynomials, it follows that
the inner product is degenerate on anti-invariants.
However, for integer~$g$, this contradicts the non-degeneracy of $(\cdot,\cdot)$ 
and the equivariance of the Dunkl operators $\mathcal D_i$ (see~\cite{DJO}).

The intertwining relation \eqref{shift} means that $K(g)$ maps eigenfunctions
of $L(g{+}1)$ to eigenfunctions of $L(g)$. By the above remark,
eigenfunctions cannot be lost. Consequently, the operator $K(g)$ maps isomorphically 
the space spanned by deformed harmonic polynomials $h_k$ with fixed $m$ and 
coupling $g{+}1$ to the space spanned by deformed harmonic polynomials~\eqref{defharm} 
with the same $m$ and coupling~$g$, and for any integer~$g$ these spaces have the 
same dimension.

Finally, the eigenstates~$\tiv_k$ of the relative spherical Hamiltonian~$\tiH_\Om$
are constructed by combining the methods outlined above. For
\begin{equation}
k=(0,0,k_3,k_4,\ldots,k_n) \quad\with 3k_3+4k_4+\ldots+nk_n=m\ ,
\end{equation}
the relative energy eigenfunctions \ $\tipsi_k=\tir^{-\frac{n-2}{2}}\tiu_k$ \
are expanded analogously to~(\ref{uexp}), dressing all equations with tildes
and shifting $h$ by~$\sfrac12$. The ground state of this system is
\begin{equation}
\tiv_0(\th) \ \propto\ \widetilde{\D}^g \qquad{\rm where}\qquad
\widetilde{\D}(\widetilde{\th}) \= \tir^{-\frac12n(n-1)} \D \=
(1-n\widehat{X}^2)^{-\frac14n(n-1)} \widehat\D(\th)\ .
\end{equation}
Note that $\widetilde\th\neq\th$.
For small values of~$n$, the function $\widetilde{\D}=\widetilde{\D}_n$ is readily computed:
\begin{equation}
\widetilde{\D}_2=1\ ,\qquad
\widetilde{\D}_3=\cos 3\widetilde\phi\ , \qquad
\widetilde{\D}_4=\sin^2\!\widetilde\th\cos^4\!\widetilde\th\cos2\widetilde\phi
(\cos^2\!\widetilde\phi\tan^2\!\widetilde\th{-}1)
(\sin^2\!\widetilde\phi\tan^2\!\widetilde\th{-}1)\ .
\end{equation}
The excited state wave functions may be obtained in the same fashion,
either by projecting first to the center-of-mass ground state and then to the
relative radial oscillator ground state,
or by employing Dunkl operators, after projecting the Newton sums $\sum_i\cD_i^\ell$
onto the hyperplane given by $\sum_i x^i=0$ and omitting any power of the first
Newton sum.
The degeneracies are further reduced in accord with the equations already presented.

\bigskip

\section{Generalized angular models}

\noindent
Let $\mathcal R \subset \R^n$ be a Coxeter root system \cite{Humphreys} with associated finite reflection group $W$. Let $g_\alpha=g(\alpha)\ge0$ be a $W$-invariant multiplicity function on the set of roots $\alpha=(\alpha_1, \ldots, \alpha_n) \in \mathcal R$. Consider the corresponding Calogero--Moser Hamiltonian \cite{olpe2} (see also \cite{Wolfes,Olsha} for rank two)
\begin{equation} \label{qHCox}
H \= \sfrac12 \sum_i p_i^2\ +\ \sfrac{\o^2}{2}\sum_i (x^i)^2\ +\
\sum_{\alpha\in {\mathcal R}_+} \frac{g_\alpha(g_\alpha{-}1)(\alpha\cdot \alpha)}{2(\alpha \cdot x)^2}\ , 
\end{equation}
where ${\mathcal R}_+$ is a positive half of the root system $\mathcal R$ and $\cdot$ is the standard scalar product in $\R^n$. Our previous analysis can be extended to the corresponding angular Hamiltonian $H_\Omega$ given by \eqref{Homrep}. 

Let $d_1{=}2, d_2, \ldots, d_n$ be the degrees of basic homogeneous $W$-invariant polynomials $\sigma_1{=}r^2, \sigma_2, \ldots, \sigma_n$. Then the spectrum of the Hamiltonian \eqref{qHCox} is given by \cite{olpe2}
\begin{equation} \label{ECox}
E_k \= \o\,\Bigl(\sum_{\alpha \in {\mathcal R}_+} g_\alpha +\sfrac{n}{2}+\sum_{i=1}^n d_i k_i\Bigr)
\quad\with k_i=0,1,2,\ldots\ .
\end{equation}
The spectrum of $H_\Omega$ has the form
\begin{equation}\label{veCox}
\ve_k\= \sfrac12 q\,(q+n-2) \quad\with\quad
q\= \sum_{\alpha \in {\mathcal R}_+} g_\alpha +m \und m\=\sum_{i=2}^n d_i k_i\ .
\end{equation}
The corresponding eigenfunctions are given by $r^{-q}\Delta^g h_k(x)$, where 
\begin{eqnarray}
\Delta^g &=& \prod_{\alpha \in {\mathcal R}_+} (\alpha\cdot x)^{g_\alpha}\ , \\
h_k(x) &=& r^{2(q-1)+n}\,\sigma_2(\cD_i)^{k_2} \sigma_3(\cD_i)^{k_3}
\cdots \sigma_n(\cD_i)^{k_n}\,r^{2(m-q+1)-n}\ ,\\[8pt]
\label{dunklCox}
\cD_i &=& \pa_i\ +\sum_{\alpha \in {\mathcal R}_+} \frac{g_\alpha \alpha_i}{\alpha \cdot x} 
\,(1-s_{\alpha})\ ,
\end{eqnarray}
and $s_\alpha$ is the orthogonal reflection about $\alpha\cdot x =0$ (cf.~\cite{dunkl}). 

Let $S\subset \mathcal R$ be $W$-invariant. Consider the differential operator $K$ which has the form
\begin{equation}
K\=\Bigl[\prod_{\alpha \in S\cap {\mathcal R}_+}\!\!\alpha\cdot\cD\Bigr]  
\Bigl[\prod_{\alpha \in S\cap {\mathcal R}_+}\!\!\alpha\cdot x\,\Bigr]\ ,
\end{equation}
with $\cD=(\cD_1,\ldots, \cD_n)$ when acting on $W$-invariant functions. Let $1_S$ be a multiplicity function which takes value 1 on the roots from $S$ and is zero otherwise. Then $K$ satisfies the intertwining relation \cite{Heckman} 
\begin{equation}
L(g)\,K \= K\;L(g{+}1_s)\ ,
\end{equation}
where $L(g)$ is the Calogero--Moser operator in the potential-free gauge: 
\begin{equation}
L(g)\= \sum_i \pa_i^2  +
\sum_{\alpha\in{\mathcal R}_+}\frac{2g_\alpha}{\alpha\cdot x}\,(\alpha\cdot\partial)\ .
\end{equation}
Similarly to the type $A$ analysis above, the operator $K$ yields an isomorphism of the spaces spanned by deformed harmonic polynomials $h_k$ at couplings $g$ and $g{+}1_S$, and the degeneracies of the spectra of the operators $H_\Omega$ at integer couplings coincide.

Another approach to construct eigenfunctions is to use  the exchange-creation operators and to separate the radial component. These operators have the form \eqref{excha} where $\cD_i$ are now given by \eqref{dunklCox}, and they satisfy the commutation relations
\begin{equation}
[a_i^\dagger, a_j^\dagger]\=[a_i, a_j]\=0 \und
[a_i^\dagger, a_j]\= -2 \omega \delta_{ij} - 4\omega \sum_{\alpha\in {\mathcal R}_+}\frac{\alpha_i \alpha_j g_\alpha}{\alpha\cdot \alpha} s_\alpha\ .
\end{equation}
Therefore, on $W$-invariant functions, we have the operator coincidence
\begin{equation}
{\mathcal H}\ :=\ 
\sum_{i=1}^n a_i^\dagger a_i\ +\ 2 \omega \sum_{\alpha\in {\mathcal R}_+} g_\alpha s_\alpha
\= -L(g)+\omega^2 r^2 -\omega n\ .
\end{equation}
This operator obeys the commutation relations (cf.~\cite{Gol}, \cite{spinCal})
\begin{equation}
[{\mathcal H}, a_j^\dagger] \= 2 \omega a_j^\dagger \und
[{\mathcal H}, a_j] \= -2 \omega a_j\ .
\end{equation}
Application of $W$-invariant combinations of the creation operators $a_j^\dagger$ to the ground state leads to the energy eigenstates. 

\bigskip

\section{Angular spin--Calogero model}

\noindent
The Calogero model can be extended to a model with particles carrying
internal degrees of freedom (which we  shall call ``spin", although
there is no notion of space  rotations under which they transform)
and interact through ferromagnetic or antiferromagnetic interactions.
By reducing this model to radial and angular variables we can recover
angular spin--Calogero models and obtain their wavefunctions.

The most straightforward way to obtain the spin--Calogero
models is to start with the full exchange-Calogero model
(not projected to spatially symmetric or antisymmetric
states) and endow the particles with $s$ internal states
(spin), which we will view as being in the fundamental representation
of SU$(s)$. The fundamental single-particle SU$(s)$ generators
${\vec S}_i = \{ S_i^a \}$ ($i=1,\dots,n$, $a=1,\dots,s^2{-}1$)
acting on the states of particle $i$ span the set of all possible spin operators.

So far, spins do not participate in the dynamics, as the Hamiltonian
does not involve any ${\vec S}_i$ and has a trivial symmetry SU$(s)^n$.
We now restrict our states to be totally symmetric (bosonic) or antisymmetric
(fermionic) under full particle exchange, swapping both space and
internal degrees of freedom. Calling $s_{ij}$ and $\sigma_{ij}$ the
corresponding space and spin exchange operators, the total permutation
operator is $P_{ij} = s_{ij} \sigma_{ij}$, and we impose
\begin{equation}
P_{ij} | \psi \rangle \= s_{ij} \sigma_{ij} | \psi \> \= \pm | \psi \>\ .
\end{equation}
So on bosonic (fermionic) states $s_{ij} = \pm \sigma_{ij}$. Replacing
all $s_{ij}$ with $\pm\sigma_{ij}$ in the exchange-Calogero Hamiltonian we
obtain the spin model
\begin{equation}
H \= \sfrac12 \sum_i p_i^2\ +\ \sfrac{\o^2}{2}\sum_i (x^i)^2\ +\
\sum_{i<j} \frac{g(g{\mp}\sigma_{ij})}{(x^i{-}x^j)^2}\ .
\end{equation}
On fundamental SU$(s)$ states the spin exchange operator can be expressed as
\begin{equation}
\sigma_{ij} \= \half \bigl( {\vec S}_i \cdot {\vec S}_j + \sfrac{1}{s} \bigr)\ .
\end{equation}
Hence, the Hamiltonian includes a (position-dependent) two-body spin coupling.
Bosonic (fermionic) states give an (anti)ferromagnetic interaction respectively.
The original SU$(s)^n$ symmetry is reduced to a global SU$(s)$, corresponding
to ${\vec S} = \sum_i {\vec S}_i$, when restricted to the above states.

The reduction of the above model to radial and angular parts proceeds exactly
as in the spinless model. The angular Hamiltonian $H_\Om$ involves the corresponding
spin interactions, while the radial part remains the same.

The spin--Calogero model has the same energy eigenvalues as the spinless model
(as can be seen by applying exchange-creation operators on the ground state)
but with different degeneracies. Each energy level can be characterized by
its spin representation content, which is a direct sum of SU$(s)$ irreps.
These spin degeneracies are the same as in the free ($g${=}0) theory. Although
their many-body state reconstruction is quite simple, their expression
in terms of the bosonized quantum numbers $k_i$ is a bit awkward.

For the bosonic model, the ground state is fully symmetric under both $s_{ij}$
and $\sigma_{ij}$.
Denoting by $[\ell]$ the $\ell$-fold symmetric irrep of SU$(s)$ (with a single
Young tableau row of length $\ell$), the ground state carries the spin state
$[n]$. The spin content of the energy level $E_k$ corresponding to
$k=(k_1 , k_2 , \dots, k_n)$ can be expressed as
\begin{equation}
[c_1 ] \times [c_2{-} c_1 ] \times \cdots \times [c_r{-} c_{r-1} ]
\times [n -c_r ]
\end{equation}
where $c_1 < c_2 < \dots < c_r$ are the indices (positions) $i$
of all {\it nonvanishing\/} $k_i$.
For example, $k=(1,0,2)$ has $r{=}2$ nonvanishing entries, in the positions
$c_1=1$ and $c_2=3$. The above can be decomposed
into SU$(s)$ irreps using standard Young tableau techniques.

The states of the antiferromagnetic spin model are harder to express,
as they depend nontrivially on the spin of the vacuum. Denoting by
$\{ \ell \}$ the $\ell$-fold antisymmetric irrep of SU$(s)$ (with a single
Young tableau column of length $\ell$), the vacuum carries the spin state
$\{ n\ {\rm mod} (s) \}$. The expression for excited states in terms of
the $k_i$ are quite complicated and will not be reproduced here.

The corresponding spin states for the angular spin--Calogero model can be
obtained in two ways: either by direct construction or by reduction from the
full spin--Calogero model, as in the spinless case. We will present both
methods and give the final result.

The construction of angular spin--Calogero states can be done similarly to
the spinless case. Specifically, we can act on the ground state with any
polynomial of the $\cD_i$ and create (the polynomial part of) excited states.
The difference now is that these polynomials need {\it not\/} be symmetric
in the $\cD_i$ as we do not consider space-symmetric states. Instead, the
full state, involving also spin degrees of freedom, must be subsequently
(anti)symmetrized to give the corresponding bosonic (fermionic) spin state.
We will, again, analyze the bosonic case.

Consider the monomial
\begin{equation}
\cD_{i_1}^{b_1} \cD_{i_2}^{b_2} \cdots \cD_{i_n}^{b_n}
\label{monom}
\end{equation}
acting on the ground state, where $i_1 , \dots,i_n$ are all distinct,
and $b_i$ are defined in terms of the $k_i$ via
\begin{equation}
b_i \= \sum_{m=i}^n k_m\ .
\end{equation}
It is clear that the total degree is $m=k_1 + 2 k_2 + \dots + n k_n$.
The $b_i$ obey $b_i \ge b_{i+1}$ and are, essentially, bosonic single-particle
excitation numbers. The total state has, now, to be symmetrized in positions
and spins combined. Any particle index $i$ appearing in a {\it unique\/} power
of $\cD_i$ transforms in the fundamental of the symmetric group S$_n$ under
permutations and, combined with the spin states, will give a fundamental
of SU$(s)$. Two indices $i_a$ and $i_{a+1}$ appearing with equal powers of $\cD$
(i.e., with $b_a = b_{a+1}$), however, are already symmetric and, therefore,
transform as the two-symmetric irrep of the symmetric group under permutations.
Combined with spin states they will give the $[2]$ irrep of SU$(s)$.
Similarly, for $b_a = \dots = b_{a+\ell -1}$ we will obtain, upon symmetrization,
the $[\ell ]$ irrep of SU$(s)$, and so on for all indices.
The remaining spin states, corresponding to indices that do not
appear in the above monomial (i.e., for vanishing $b_i$) will be symmetrized
among themselves and will produce $[n{-}r]$, where $r$ is the number of nonvanishing
$b_i$. The total spin state is the direct product of all the above symmetric states.

It is easy to see that this construction reproduces the {\it full\/} set of spin states
of the spin--Calogero model as described previously.\footnote{
Note that equality of successive $b_i$ means vanishing of the corresponding $k_i$.}
What is missed in the above analysis is the possibility that some of such states
may vanish. And they do.
As an example, consider $\cD_i^2$. Acting on the ground state would produce
\begin{equation}
[1] \times [n{-}1] \=  [n] \ +\  [n{-}1 , 1]
\end{equation}
where $[n{-}1,1]$ refers to the Young tableau with rows of length $n{-}1$ and $1$.
Among these symmetrized states, however, are states where the index $i$ is
symmetrized separately, corresponding to $\sum_i \cD_i^2$.
Those states would contribute the irrep $[n]$ in the spin states,
but they actually vanish. Hence, the true spin state built from $\cD_i^2$ is just
$[n{-}1,1]$.
So the task of constructing all spin states of the angular model amounts to
identifying all components that would be produced from the subset of states
involving (powers of) $\sum_i \cD_i^2$ and eliminating them. Clearly the set of
all such states at level $m$ can be obtained by acting with the operator
$\sum_i \cD_i^2$ times monomials of the form (\ref{monom}) of degree $m{-}2$.
These states span the representation content of spin--Calogero states at level
$m{-}2$, since $\sum_i \cD_i^2$ is a singlet. Removing these states from the
full set of states we obtain the simple and explicit result
\begin{equation}
S_\Om (m) \= S(m) - S(m{-}2)
\label{spinang}
\end{equation}
where $S_\Om (m)$ is the SU$(s)$ representation content of the angular spin--Calogero
model at level $m$ while $S(m)$ is the SU$(s)$ representation content of the full
spin--Calogero model. This is analogous to the corresponding formula~(\ref{degsym})
for the degeneracies of the spinless model, where all spin ``irreps'' are the singlet.
It is easy to see that, e.g., the above formula reproduces the reduction at level two
due to $\cD_i^2$ that we discussed above.

The same result is obtained using the reduction method.
The states of the full spin--Calogero model can be produced by combining the
states of the angular Hamiltonian with the radial part. The spectrum and degeneracy
of the radial Hamiltonian are exactly the same as in the spinless case for
each eigenvalue of the angular Hamiltonian, while the full spin content
of the states is provided by the angular states. An analysis similar to the
spinless case yields again the result (\ref{spinang}).
 The first few spin states are listed below:
\begin{equation}
\begin{aligned}
S_\Om (0) &\= [n] \ ,\\
S_\Om (1) &\= [1] \times [n{-}1] \ ,\\
S_\Om (2) &\= [2] \times [n{-}2]\ +\ [1] \times [n{-}1]\ -\ [n] \ ,\\
S_\Om (3) &\= [1] \times [1] \times [n{-}2]\ +\ [3] \times [n{-}3] \ ,\\
S_\Om (4) &\= [1] \times [1] \times [n{-}2]\ +\ [1] \times [2] \times [n{-}3]
\ +\ [4] \times [n{-}4]\ .
\end{aligned}
\end{equation}

Finally, we can produce a relative angular spin--Calogero model by separating the
center-of-mass degrees of freedom from the full spin--Calogero model and then
reducing to radial and angular parts, as in the spinless case. The analysis of the
spectrum and states is similar to the one for the angular spin--Calgero model above
and will not be repeated here. Using either direct construction or the reduction
method, we obtain the representation content of states at level $m$ for the
bosonic (ferromagnetic) case as
\begin{equation}
{\widetilde S}_\Om(m) \= S_\Om(m) - S_\Om(m{-}1) \= S(m) - S(m{-}1) - S(m{-}2) + S(m{-}3)
\end{equation}
again a result similar to (\ref{alldeg}) for the spinless case.

We conclude by mentioning that the spin--Calogero model can yield a corresponding 
spin chain model using the ``freezing" trick: by driving the coupling constant~$g$ 
and frequency~$\omega$ to infinity, the coordinates ``freeze'' in their classical 
equilibrium configuration and the spin degrees of freedom decouple~\cite{spinchain}. 
The angular (reduced or not) spin-Calogero model in that limit gives rise to the 
same spin chain model, since the radial and center-of-mass coordinates belong 
to the decoupling kinematical degrees of freedom, and the angular Hamiltonian
captures the full spin content of the model.  

\bigskip

\noindent
{\bf Acknowledgments}
\nopagebreak

\noindent
We thank Tigran Hakobyan and Armen Nersessian for discussions.
This work was partially supported by 
the Royal Society/RFBR joint project JP101196/11-01-92612, by
the Volkswagen Foundation under grant I/84~496,
by the Deutsche Forschungsgemeinschaft under grant LE 838/12-1, 
by the National Science Foundation under grant PHY/1213380 and by the Research
Foundation of CUNY under grant PSC-CUNY 65501-0043.

\bigskip

\end{document}